# Authentication Systems in Internet of Things

**\*Tuhin Borgohain**
Department of Instrumentation Engineering, Assam Engineering College, Guwahati, India
Email: borgohain.tuhin@gmail.com
*Corresponding author
**Amardeep Borgohain**
Department of Electrical Engineering, Assam Engineering College, Guwahati, India
Email: amardeepborgohain@gmail.com
**Uday Kumar**
Tech Mahindra Limited, Chennai, India
Email : udaykumar@techmahindra.com
**Sugata Sanyal**
Corporate Technology Office, Tata Consultancy Services, Mumbai, India
Email: sugata.sanyal@tcs.com

------------------------------------------------------------------ABSTRACT-------------------------------------------------------
**This paper analyzes the various authentication systems implemented for enhanced security and private reposition of an individual's login credentials. The first part of the paper describes the multi-factor authentication (MFA) systems, which, though not applicable to the field of Internet of Things, provides great security to a user's credentials. MFA is followed by a brief description of the working mechanism of interaction of third party clients with private resources over the OAuth protocol framework and a study of the delegation based authentication system in IP-based IoT.**

Keywords – **Authentication, authorization, OAuth, SASL**

## I. INTRODUCTION

Wide adoption of seamless integration of social network with any electronic devices like a smart phone, fitness tracker etc. has facilitated the interaction of personnel with anyone over the world with just the click of a button. This has paved the way for sharing of personal health-related data [1], online social account and many more with anyone in any continent and has opened the way for social collaboration and competition over the Internet. Yet the communication taking place between the smart objects and the social media is susceptible to interception from various third party intruders, which may lead to loss of privacy and exposure of authentication details to unwanted personnel. Thus an improved and more robust authentication system from logging inside one's online account will ensure a much safer browsing experience and online exchange of information.

## II. OVERVIEW

In section III, we will briefly introduce the authentication system in a social network. In section IV, we will discuss the two-step authentication [4] system adopted by many online tech services and non-IoT devices to decrease the chance of logging inside someone's account by another person due to accidental compromise of password. In section V, we discuss about the OAuth framework applicable in transmission of data in IoT devices. In section VI, we discuss the delegation-based authentication method for IP-based IoT. In section VII, we conclude the paper.

## III. AUTHENTICATION

In any social network connected to a smart object, a link between the online portal and the Internet connected device is initiated by the upload of the registered user ID and its corresponding password to an online server. Authentication is the process of confirming one's identity. Two-factor authentication or 2FA is a way of login where a user is required to provide additional information to sign in than just the password. Using only the password to login enables malicious attackers to have easy access into the system as it represents single piece of information only [14]. In authentication systems, all the transmission of data from a user's smart object to the online server can be exposed to unwanted personnel through interception. As such, from a security standpoint, the most common authentication systems fail to guarantee a fail-safe method for keeping the login information away from the hands of the public for maintaining privacy and security for the user. With the increasing use of mobile devices by the consumers for banking, shopping [17] etc. the need for security concerns have emerged which in turn has created an interest in multi-factor authentication. In case of multi-factor authentication, which requires more than one form of authentication for verification of legitimacy, provides an additional layer of protection against security breach. Here apart from providing username and password by the user, an additional authentication code is sent to the user's mobile device [13]
for verification. These factors taken together provide increased security of accounts. Yet such multi factor authentication systems are not applicable in the Internet of Things architecture.

## IV. TWO-STEP AUTHENTICATION IN NON-IoT DEVICES

In some of the leading tech companies offering online services, the companies offer the choice of Multi-factor authentication system for enhanced security. In this type of authentication system, verification of specific combination of multiple components takes place before authenticating an individual. If one of the components is missing/incorrect, the identity of the individual is not accepted as registered in the system. For enhancing the security against intruders, while transferring data, data hiding technique [8] can be implemented by the companies in order to prevent data theft.

One of the most prominent multi-factor authentication systems is the mobile phone two-factor authentication [2]. In this system, a user accesses his account by inputting the registered information along with a dynamic one-time valid password [5] made up of combination of digits. The latter password is sent to the user through email, SMS or any other specific application designed to facilitate the procurement of the OTP (One Time Password) ([9], [10], [11], [22], [30]) from the host server. To ensure the security of the OTP, its validity is limited to a very short duration of time [7] beyond which the OTP expires and the user has to re-enter the personal information corresponding to which a new OTP is requested, generated and sent to the user from the server. A basic flow diagram of how MPTFA works is shown in fig 1:

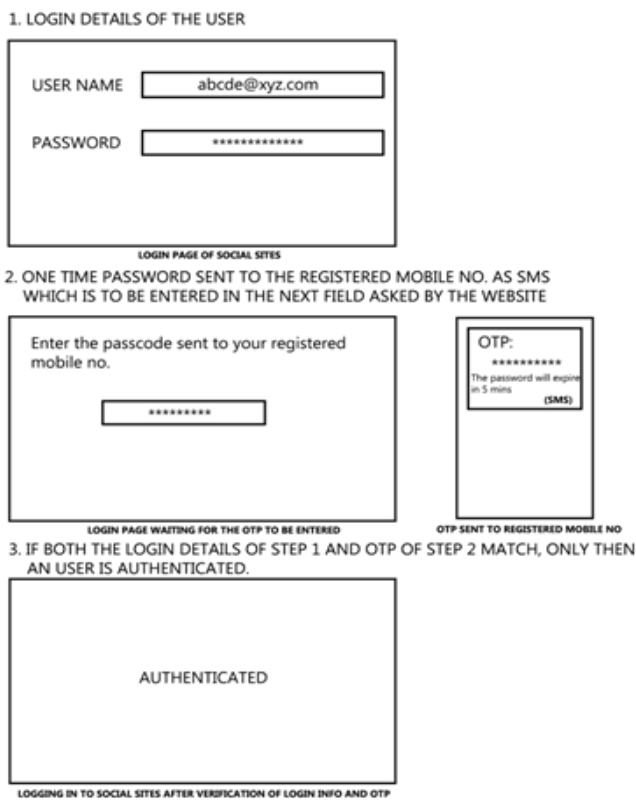

Fig 1: Flow diagram of MPTFA

Some of the major advantages of the mobile phone two-factor authentication (MPTFA) over other multi-factor authentication systems are:
i) As against the requirement of tokens like USB, bank cards etc. by the user all the time for verifying their identity, MPTFA replace the tokens by an individual's mobile phone which is carried out the entire time user.
ii) The ability to limit the maximum number of permitted false entries in MPTFA reduces the chance of the information being compromised through hit-and-trial methods.
iii) The OTP are dynamically generated which changes with each entry of the static login details which proves to be more secure than only the static login information.
iv) The short expiration duration of the passwords ensures that an OTP left in the mobile phone cannot be used again by anyone to gain entry into an account in the future.

Some drawbacks of the MPTFA are as follows:
i) The mobile phone's capability of receiving messages depends on its cellular reception [15] and battery longevity.
ii) The loss of a mobile phone containing the SIM card registered for receiving the OTP leads to additional distress for the user as the absence of the mobile phone do not let the user complete his login steps which leads to failure in accessing his/her own account.
iii) From an economic viewpoint, the replacement of a lost mobile phone by another proves to be an expensive affair for the common man.
iv) In terms of privacy, the user has to share his personal telephone number with the provider, which means lack of privacy [20] up to a certain extent as the provider may choose to disclose the number to the public or some other advertising firms.

## V. AUTHENTICATION USING OAUTH FRAMEWORK OVER SIMPLE AUTHENTICATION AND SECURITY LAYER (SASL) IN IoT DEVICES

OAuth is an open standard authorization and authentication protocol framework granting the third-party applications a limited, delegated access to private resources by establishing an approval interaction between the third-party application and the resource owner and specifying a definite process by which the resource owner grants authorization to the third-party applications access to the server resources without giving away their login information (User ID, passwords etc.) ([18], [21], [23], [26], [27]).

On the other hand, simple layer authentication and security layer (SASL) is an authentication framework for data security in application layer framework ([19], [25], [28], [29]).

During grant of access to client (say Facebook app, Twitter app etc.) to protected resources (User's Facebook account, Twitter account etc.), initially the requests for permission to resource access in done over Plain OAuth 2.0 but the last stage for authenticating a client to access the resources from a resource owner is done using the OAuth protocol framework over the SASL authentication framework [12]. The systematic transmissions of requests from the client for grant of authorization are given below:

Using Plain OAuth 2.0
Step I: The client request for a grant of authorization from the resource owner in two ways:
i) Resource owner gets the request sent by client, directly.
ii) The request is sent through an intermediate authorization server by the client.

Step II: An authorization is granted to the client in the form of a credential. This authorization depends on the whether the client requested for the grant directly or indirectly.

Step III: The resource server can only be accessed with specific access token. These are requested by the client by first authenticating themselves with the authorization server and then forwarding the grant of authorization received directly from the resource owner or indirectly via the authorization server.

Step IV: If the client is an authenticated in their servers, the authorization server verifies the authorization grant and then issues an access token.

Using OAuth over SASL:

Step I: After gaining the access token, the client requests for access to the private resources form the resource server by authenticating themselves with the access token.

Step II: The resource server verifies the access token. If successful, the client is authenticated to access the resources on behalf of the resource owner.

All the above steps are represented diagrammatically in the following Fig 2.

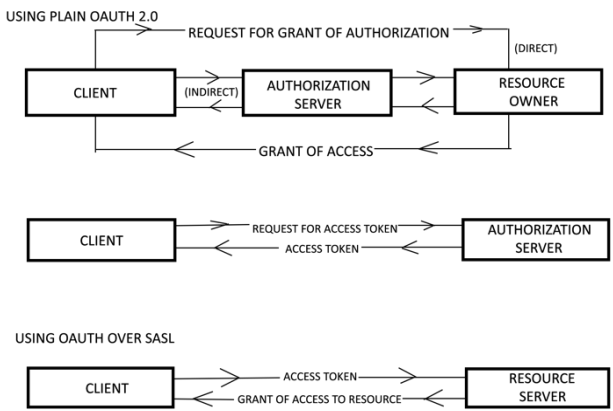

Fig 2: Communication between client and resource server over OAuth framework

## VI. DELEGATION BASED DTLS CONNECTION ESTABLISHMENT FOR AUTHENTICATION IN IoT

This is an authentication method for Internet of Things proposed by Hummen *et al.* [31].

DTLS stands for Datagram Transport Layer Security. In a delegated DTLS connection establishment, a delegation server is used for separating the initial establishment of connection from the subsequent application data protection. Here the delegation server makes the provision for a constrained device with all the necessary contexts of security for secured communication. These delegation servers do not contain any pre-shared secret keys for the communicators but establishes security contexts on-demand.

First a master key in imprinted in the constrained device by the delegation server for transmission of security contexts to and fro between the devices. This imprinting is done when the delegation server is bootstrapped to the local network domain. Under the assumption of administration by a common operator, the inter-connection between the constrained device and a remote end-point is established by the instruction of the operator to the delegation server to establish a DTLS connection with the aforementioned end-point. During this connection between the delegation server and the remote end-point, the former authenticate the latter during the DTLS handshake. This authentication is done via trusted certificates. To achieve the goal of handing over the security context to the constrained device form the above established connection, the session resumption extension of DTLS protocol is employed during the establishment of the above connection. This session resumption extension of DTLS protocol facilitates two things:

(i) Sufficient information for re-establishment of connection between the delegation server and the remote end-point even after the terminal of the connection between the two is maintained.

(ii) It allows a secured transfer of security contexts between the two end-points through an encrypted session ticket.

The delegation server uses the above mechanism to for transfer of security contexts to the remote end-point. Through this transfer, the delegation server encrypts the transferable security contexts with the master key pre-imprinted in the constrained device. Moreover, during this handshake between the delegation server and the remote end-point, the server makes an attachment of the constrained device's IP address to the session ticket.

After the handshake is completed, the remote endpoint initiates a *session resumption handshake* with the constrained device through the previously attached IP-address. Similarly, *session resumption handshake* with the constrained device is triggered by the delegation server through the IP address of the remote end-point. During these *session resumption handshake* transfer of the session ticket along with the security context takes place from the remote end-point to the constrained device where decryption of the security context takes place. These security contexts are then used for authentication and re-establishment of connection shared previously with the remote end-points.

## VII. CONCLUSION

The above authentication methods in Internet of Things has resulted in more secure login experience for the users without the dangers of exposing their credentials to third party users. Further research for improvement of authentication systems in IoT will result in a much wider adoption of IoT in daily life along with ensuring greater privacy and security on part of the user during the login processes and financial transactions over the net. Moreover efficient transmission of data after securing the transmission process ([3], [6]) is of utmost importance, which can be obtained by Ant Colony Optimization (ACO) algorithm ([16], [24]).

**Biographies and Photographs**

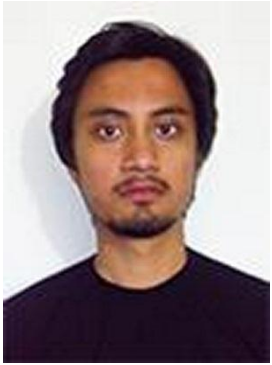
Tuhin Borgohain is a 3rd Year student of Assam Engineering College, Guwahati. He is presently pursuing his Bachelor of Engineering degree in the department of Instrumentation Engineering.

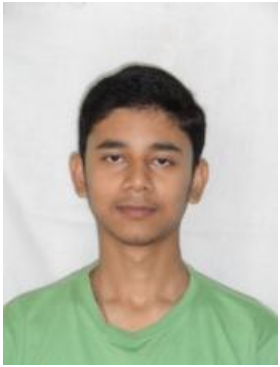
Amardeep Borgohain is a 3rd Year student of Assam Engineering College, Guwahati. He is presently pursuing his Bachelor of Engineering degree in the department of Electrical Engineering.

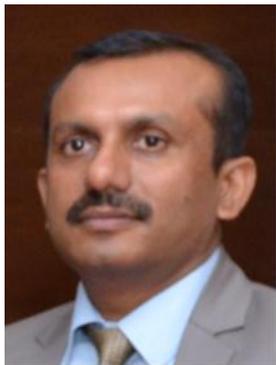
Uday Kumar is working as Delivery Manager at Tech Mahindra Ltd, India. He has 17 years of experience in engineering large complex software system for customers like Citibank, FIFA, Apple Smart objects and AT&T. He has developed products in BI, performance testing, compilers. And have successfully led projects in finance, content management and ecommerce domain. He has participated in many campus connect program and conducted workshop on software security, skills improvement for industrial strength programming, evangelizing tools and methodology for secure and high end programming.

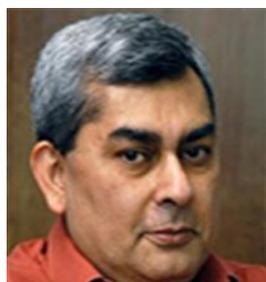
Sugata Sanyal is presently acting as a Research Advisor to the Corporate Technology Office, Tata Consultancy Services, India. He was with the Tata Institute of Fundamental Research till July, 2012. Prof. Sanyal is a: Distinguished Scientific Consultant to the International Research Group: Study of Intelligence of Biological and Artificial Complex System, Bucharest, Romania; Member, "Brain Trust," an advisory group to faculty members at the School of Computing and Informatics, University of Louisiana at Lafayette's Ray P. Authement College of Sciences, USA; an honorary professor in IIT Guwahati and Member, Senate, Indian Institute of Guwahati, India. Prof. Sanyal has published many research papers in international journals and in International Conferences worldwide: topics ranging from network security to intrusion detection system and more.